\documentclass[11pt]{article}
\usepackage{amsfonts,amssymb,amsmath}
\begin{document}
\renewcommand{\theequation}{\thesection.\arabic{equation}}
\newcommand{\bea}{\begin{eqnarray}} 
\newcommand{\eea}{\end{eqnarray}}
\newcommand{\RR}{{\mathbb R}}
\newcommand{\NN}{{\mathbb N}}
\newcommand{\ZZ}{{\mathbb Z}}
\newcommand{\CC}{{\mathbb C}} 
\renewcommand{\SS}{{\mathbb S}}
\newcommand{\vac}[1]{\langle #1\rangle}
\newcommand{\norm}[1]{\vert\!\vert #1\vert\!\vert}
\newcommand{\wick}[1]{{:}#1{:}}
\newcommand{\R}[2]{{\bf Result #1}: {\sl #2}}
\renewcommand{\today}{}

\title{\bf Pole structure and biharmonic fields in conformal QFT in
  four dimensions\footnote{Lecture given by K.-H.R. at the Workshop
    ``Lie Theory and Its Applications in Physics'', 18--24 June 2007,
    Varna, Bulgaria; to appear in the proceedings}} 

\author{Nikolay M.\ Nikolov$^{1}$, \\[2mm]
Karl-Henning Rehren$^{1,2}$, \\[2mm]
Ivan Todorov$^{1}$}

\maketitle

\begin{center}
\vskip-10mm
\scriptsize
\parbox{300pt}{
\begin{itemize}
\item [$^{1}$]
Institute for Nuclear Research and Nuclear Energy, \\
Tsarigradsko Chaussee 72, BG-1784 Sofia, Bulgaria; \\
mitov@inrne.bas.bg, todorov@inrne.bas.bg
\item [$^{2}$]
Institut f\"ur Theoretische Physik, Universit\"at G\"ottingen, \\
Friedrich-Hund-Platz 1, D-37077 G\"ottingen, Germany; \\
rehren@theorie.physik.uni-goe.de
\end{itemize}}
\end{center}

\begin{abstract}
Imposing Huygens' Principle in a 4D Wightman QFT puts strong
constraints on its algebraic and analytic structure. These are best
understood in terms of ``biharmonic fields'', whose properties
reflect the presence of infinitely many conserved tensor currents. 
In particular, a universal third-order partial differential equation
is derived for the most singular parts of connected scalar correlation
functions. This PDE gives rise to novel restrictions on the pole
structure of higher correlation functions.
An example of a six-point function is presented that cannot arise from
free fields. This example is exploited to study the locality
properties of biharmonic fields. 
\end{abstract}

\section{Introduction}
\setcounter{equation}{0} 

In spite of the quantitative successes of renormalized perturbation
theory, no nontrivial quantum field theory (QFT) in four spacetime
dimensions (4D) has been constructed rigorously. It is sometimes
suggested that the Wightman axioms might be too restrictive, but most
attempts at relaxing them lead to physically unacceptable consequences.
E.g., a violation of causality cannot be exponentially small 
\cite{P}. Admitting indefinite (physical) Hilbert spaces not only
jeopardizes the statistical interpretation of correlations, but it 
makes statements about convergence of approximations delicate if not
meaningless.   

On the contrary, the usual attitude is to strengthen the axioms, e.g., by
imposing additional symmetries or phase space or additivity properties. 
This facilitates the analysis of models but clearly reduces the number of
theories. Even worse, demanding too much may trivialize the theory. 
E.g., requiring that the two-point function be supported on a mass
shell \cite{JS} or only to have finite mass support \cite{G}, already
forces the field to be a (generalized) free field.   

It is one of the merits of the axiomatic approach \cite{SW,FRS} that it
allows to pinpoint such 
obstructions even before a model is formulated. It makes no
assumptions about intermediate steps and limits through which a theory
is constructed. Referring only to its intrinsic features, it avoids
assigning significance to the artifacts of the description. 
Especially, it guides navigation between Scylla (physical meaningless
theories) and Charybdis (free fields) in the quest for a proper set of
model assumptions that can possibly be satisfied by other than free fields. 

\smallskip

In a series of recent papers \cite{NST,NT}, two of us
have suggested to demand ``global conformal invariance'' (GCI) which
is the postulate that the conformal group is represented in a true
(i.e., not a covering) representation. (The term ``global'' stresses
the fact that the conformal transformations are defined globally on
the Dirac compactification of Minkowski spacetime.)  

GCI implies rationality of all correlation functions \cite{NT}. On
one side, this is a desired feature since it allows to
parameterize each correlation function by the coefficients of a finite
set of admitted rational structures. The latter are determined by
conformal invariance, and by the unitarity bound on the representations
of the conformal group that possibly contribute to operator product
expansions (OPE) giving rise to upper bounds on the pole in each
pair of variables. The coefficients are then further restricted only
by Hilbert space positivity. While the latter is highly nontrivial to
control in general, there has been considerable progress at the four-point
level \cite{DO,NRT1}. 

On the other side, GCI is equivalent to Huygens locality \cite{NRT1},
i.e., commutativity of fields not only at spacelike but also at
timelike separation. This feature seems to be conspicuosly close to
free field theory, since any scattering of field quanta should give rise to
causal propagation within the forward lightcone. Indeed, it has been
shown \cite{B} that if a Huygens local QFT has a complete particle 
interpretation, then its scattering matrix is trivial. Therefore, GCI
allows only nontrivial theories without asymptotic completeness, which
may not be entirely unphysical in a scale invariant theory. Note that by
rationality, all scaling dimensions are integer numbers, which does
not mean that they are canonical. It is conceivable that in a QFT with
anomalous dimensions, at some value of the coupling constant all
dimensions simultaneously become integers. Examples for such a hazard
are well known in 2D conformal QFT \cite{BMT}. 

\smallskip

We shall report here on a recent analysis \cite{NRT2} of the consequences
of the following fact. For a symmetric tensor field of rank $r$ and
dimension $d$, one calls $d-r$ the ``twist''. Twist two fields are
necessarily conserved, because their two-point functions, completely
determined by $r$ and $d$, are conserved. (Hilbert space positivity
crucially enters in this argument: if the norm square of a vector
vanishes, then the vector itself vanishes. The rest of the argument
invokes the Reeh-Schlieder theorem for which in turn locality 
and energy positivity are essential.) 
These conservation laws can be
reformulated in the form 
\bea\label{bh}
\square_x V(x,y) = 0 = \square_y V(x,y)
\eea
where $V(x,y)$ is the sum of all twist two contributions to the OPE
$\phi_1(x)\phi_2(y)$ of two scalar fields of equal dimension $d$. 
We say that the ``bi-field'' $V$ is biharmonic.

Our first main result is the unravelling of a hidden consequence of this
equation: a third-order partial differential equation to be satisfied
by (certain parts of) all correlation functions involving
$\phi_1(x)\phi_2(y)$, that is a necessary and sufficient condition for
biharmonicity. 

We present a solution to this PDE, and the corresponding
transcendental six-point correlation function of $V(x,y)$ that cannot
be produced by Wick products of free fields. On the basis of this
solution, which we believe to be prototypical for the general
case, we then study the locality properties of the bi-field.

\section{Biharmonicity}
\setcounter{equation}{0} 

We first explain how harmonicity of $V(x,y)$ serves to define its
correlation functions. 
Because $V(x,y)$ is biharmonic, there are
in fact two such prescriptions, and hence $V(x,y)$ is
overdetermined. The consistency condition gives rise to a restriction
on the correlation functions involving $\phi_1(x)\phi_2(y)$.

\smallskip

The crucial fact is the following ``classical'' result \cite{BT}. If
$u(x)$ is a power series in $x\in\RR^D$ or $\CC^D$, then there is a
unique ``harmonic decomposition'' 
\bea\label{hd}
u(x) = v(x) + x^2\hat u(x),
\eea
such that $v$ is harmonic ($\square_x v=0$) and $\hat u$ is again a
power series. We call $v$ the ``harmonic part'' of $u$. (Questions
of convergence will be discussed in Sect.\ 5.) 

We apply this fact as follows. 
Twist two contributions to correlations
involving $\phi_1(x)\phi_2(y)$ have a leading singularity
$((x-y)^2)^{1-d}$, while all higher twist contributions are less
singular. Therefore, the (Huygens bilocal, but not conformally
covariant) bi-field $U(x,y)$ defined by  
\bea\label{U}
\phi_1(x)\phi_2(y) - \vac{\phi_1(x)\phi_2(y)} =: 
((x-y)^2)^{-(d-1)}\cdot U(x,y)
\eea
is regular at $(x-y)^2=0$. $U(x,y)$ contains all twist $\geq$ two
contributions to the OPE, and because the twist $>$ two contributions
are suppressed by another factor $(x-y)^2$, we may write
\bea\label{Uhd}
U(x,y) = V(x,y) + (x-y)^2\; \hat U(x,y)
\eea
where both $V$ and $\hat U$ are regular at $(x-y)^2=0$. Now consider
any correlation function
\bea\label{Ucorr}
u(x,y,\dots) =  \vac{U(x,y)\phi_3(x_3)\cdots\phi_n(x_n)},
\eea
where $\dots$ stands for the arguments of all other fields. 
Its Taylor expansion in $x$ around $y$ is a
power series in $x-y$ with coefficients independent of $x$. Thus, by
(\ref{Uhd}) and because $\square_xV(x,y)=0$, the desired
correlation function (again as a power series)
\bea\label{Vcorr}
v(x,y,\dots) = \vac{V(x,y)\phi_3(x_3)\cdots\phi_n(x_n)} 
\eea
is the harmonic part of this series. By construction,
$v(x,y,\dots)$ transforms like the correlation function of a conformal
bi-scalar of dimension $(1,1)$. 

On the other hand, the Taylor expansion of $u(x,y,\dots)$ in $y$
around $x$ is another power series in $x-y$, whose coefficients do not
depend on $y$, and because also $\square_yV(x,y)=0$, $v(x,y,\dots)$
may as well be determined as the harmonic part of this latter series. 

This overdetermination imposes a consistency condition on the function
$u(x,y,\dots)$. Its nontriviality may be seen from the following
example. Consider $u(x,y,\dots) =
(y-x_6)^2/(x-x_3)^2(y-x_4)^2(y-x_5)^2$. This function 
is harmonic with respect to $x$, hence is harmonic part with the first
prescription coincides with $u$ itself, but it is not harmonic with
respect to $y$, so the harmonic part with respect to the second 
prescription differs from $u$, and the two definitions
of $v$ are conflicting each other. Thus, a function $u(x,y,\dots)$
as in this example cannot occur as a correlation function of $U(x,y)$.

We conclude that biharmonicity of the bi-field $V(x,y)$, which
follows from the conservation of conformal twist two tensor fields,
implies a nontrivial restriction on the possible correlation functions
$u(x,y,\dots)$ of the bi-field $U(x,y)$, and hence on the
correlations involving $\phi_1(x)\phi_2(y)$. 

\smallskip

We shall now turn this condition into a partial differential equation.

Global conformal invariance implies that scalar correlation functions  
are Laurent polynomials in the variables $\rho_{ij} = \rho_{ji} =
(x_i-x_j)^2$ of the form
\bea\label{corr}
\vac{\phi_1(x_1)\cdots\phi_n(x_n)}
= \sum\nolimits_{\underline\mu} C_{\underline\mu}\;
\prod\nolimits_{i<j}\rho_{ij}^{\mu_{ij}} 
\eea
where the integer powers $\mu_{ij} = \mu_{ji}$ satisfy the homogeneity rules
\bea\label{sum}
\sum\nolimits_j \mu_{ij} = -d_i,
\eea
and the absence of non-unitary representations of the conformal group
in the OPE implies the lower bound for the connected parts of (\ref{corr}) 
\bea\label{pb}
2\mu_{ij} \geq -d_i-d_j.
\eea
Let $\phi_1$ and $\phi_2$ have the same dimension $d$. It follows
that all correlations (\ref{Ucorr}) 
(that give contributions to (\ref{corr})) are Laurent polynomials in
$\rho_{ij}$, which are separately homogeneous of total degree $-1$ in
$\rho_{1k}$ ($k\neq 1$) and in $\rho_{2k}$ ($k\neq 2$), and which are
true polynomials in $\rho_{12}$. Because all terms involving a factor
$\rho_{12}$ have zero harmonic part in the harmonic decompositions, we
need to consider only the function $u_0$ which is the contribution of
order $(\rho_{12})^0$ of $u$. Then $u_0$ is separately homogeneous of
total degree $-1$ in $\rho_{1k}$ ($k>2$) and in $\rho_{2k}$ ($k> 2$). 

It is now important that the harmonic part $v$ is a real analytic
function in a neighborhood of $x_1=x_2$, provided $(x_2-x_i)^2\neq 0$
for all $j>2$ (see Sect.\ 5). We may therefore expand
$v$ as a power series $\sum_{n=1}^\infty h_n/n! \cdot
\rho_{12}^n$. The coefficients $h_n$ are functions of all the 
remaining variables $\rho_{ij}\neq \rho_{12}$, and are separately
homogeneous of total degree $-n-1$ in $\rho_{1k}$ ($k>2$) and in
$\rho_{2k}$ ($k> 2$). 
 
Let us write
$\partial_{jk}=\partial_{kj}=\frac{\partial}{\partial\rho_{jk}}$. Then
the wave operator $\square_{x_1}$ has the form
\bea\label{wave}
\square_{x_1} =-4\Big(\sum\nolimits_{2\leq j<k\leq
  n}\rho_{jk}\partial_{1j}\partial_{1k}\Big) = -4(D_1 + E_1\,\partial_{12})\eea
valid on homogeneous functions of total degree $-1$ in $\rho_{1k}$
($k\neq 1$) \cite{NRT1}, where
\bea\label{ED}
D_1 =
\sum\nolimits_{3\leq j<k\leq n}\rho_{jk}\partial_{1j}\partial_{1k}
 \qquad\hbox{and}\qquad E_1=\sum\nolimits_{3\leq i}\rho_{2i}\partial_{1i}.
\eea
Similarly, replacing $1\leftrightarrow 2$ everywhere, one represents
$\square_{x_2} = -4(D_2 + E_2\,\partial_{12})$. Thus, the two
conditions $\square_{x_1} f = 0 =\square_{x_2} f$ give rise to two
recursive systems of partial differential equations for the
coefficient functions $h_n$ of the form 
\bea\label{rec12}
E_1 \, h_{n+1} = - D_1 \, h_n \qquad\hbox{and}\qquad
E_2 \, h_{n+1} = - D_2 \, h_n.
\eea
At $n=0$ we obtain the integrability condition  
\bea\label{int} (E_1D_2 - E_2D_1)\, h_0= (E_2E_1-E_1E_2) \, h_1.
\eea
Because $h_1$ is separately homogeneous of total degree
$-2$ in $\rho_{1k}$ ($k\geq 3$) and in
$\rho_{2k}$ ($k\geq 3$), the commutator  $(E_2E_1-E_1E_2)$ 
vanishes on $h_1$. Since $v$ is the harmonic part of $u$, its leading
term $h_0$ equals $u_0$. Thus, we arrive at 

\smallskip

\noindent \R{1}{The function $u_0$ solves the third-order partial
differential equation} 
\bea\label{pde} (E_1D_2 - E_2D_1)\, u_0= 0.
\eea

Next, when (\ref{pde}) holds and the recursion is solved for $h_1$,
one has $(D_1E_2-D_2E_1)h_1=-(D_1D_2-D_2D_1)h_0=0$ because $D_1$ and
$D_2$ commute. But $D_1E_2-D_2E_1 = E_2D_1-E_1D_2$. Therefore, the
integrability condition for the next step of the recursion is
automatically satisfied, and the argument passes to all the higher
steps. Thus, (\ref{pde}) secures solvability of the entire recursive
systems (\ref{rec12}).

\section{Consequences}
\setcounter{equation}{0} 

One should worry how there can be a new universal (model-independent)
partial differential equation for the correlation functions. It is
important to notice that this PDE can not be regarded as some
``equation of motion''. The reason is that it is satisfied only by the
leading part $u_0$ of $u$. The splitting of $u$ into $u_0+$ a
remainder does not correspond to any local decomposition of the
bi-field $U(x,y)$. Thus, the PDE (\ref{pde}) cannot be formulated as a 
differential equation for some (bi-)fields in the theory. 

Instead, it should be understood as a kinematical constraint. Because
its solutions $u_0$ must at the same time be Laurent polynomials, the
PDE rather selects a (finite) set of admissible singularity
structures, that depends on the dimensions of the scalar fields
involved through the lower bounds on $\mu_{ij}$.  

\smallskip

Indeed, we have shown in \cite{NRT2} that the PDE (\ref{pde}) implies
the following constraint on the pole structure of a Laurent polynomial
$u_0$ in $\rho_{1k}$ and $\rho_{2k}$ ($k>2$), that is homogeneous of
degree $-1$ in both sets of variables separately: Suppose $u_0$
contains a monomial  
\bea\label{mon}
\prod\nolimits_{k>2} \rho_{1k}^{\mu_{1k}} \rho_{2k}^{\mu_{2k}} \times
\hbox{other factors}
\eea
where the other factors depend only on $\rho_{kl}$ ($k,l>2$). If there
are $i\neq j$ such that $\mu_{1i}<0$ and $\mu_{1j}<0$ (a ``double
pole'' in $x_1$), then one must have $\mu_{2k}\geq 0$ for all $k>2$,
$k\neq i,j$. In particular, this excludes ``triple poles'', because a
triple pole in $x_1$ would imply that {\em all} $\mu_{2k}\geq 0$,
contradicting homogeneity. The most involved possible pole structure
of $u_0$ is therefore of the form 
\bea\label{pole}
\frac{\hbox{polynomial}}{\rho_{1i}^p\rho_{1j}^q\rho_{2i}^r\rho_{2j}^s}
\times \hbox{other factors}
\eea
where the polynomial takes care of the proper homogeneity. 
The corresponding contribution to the connected correlations involving  
$\phi_1(x)\phi_2(y)$ is therefore 
\bea\label{cont}
\frac 1{\rho_{12}^{d-1}}
\frac{\hbox{polynomial}}{\rho_{1i}^p\rho_{1j}^q\rho_{2i}^r\rho_{2j}^s}
\times  \hbox{other factors}.
\eea
Note that no constraints arise on higher twist contributions
($\mu_{12} > 1-d$) or on poles of two fields with $d_1\neq d_2$. 

\smallskip

The interest in double poles is due to the fact that twist two bi-fields
made of free fields, such as $\wick{\varphi(x)\varphi(y)}$ or
$(x-y)^\mu\wick{\bar\psi(x)\gamma_\mu\psi(y)}$ are always Wick
bilinears, so that their correlation functions can never contain a
double pole. A nontrivial double pole solution (an example will be
displayed below) is therefore a candidate for a Huygens local QFT not
generated by free fields.  

We have also shown in \cite{NRT2} that 

\smallskip

\noindent \R2{A correlation function involving
$V(x_1,x_2)$, i.e., the harmonic part of the Laurent polynomial
$u_0(x_1,x_2,\dots)$, is again a Laurent polynomial if and only if
$u_0$ does not contain a double pole in $x_1$ or $x_2$.  }

\smallskip 

Four-point functions $\langle U(x_1,x_2)\phi_3(x_3)\phi_4(x_4)\rangle$
can never exhibit double poles in $x_1$ or $x_2$, just ``by lack of
independent variables''. Therefore, four-point functions of 
twist two bi-fields are always rational. From this one can deduce
that their partial wave expansion cannot terminate after finitely many
terms, i.e., the OPE of $\phi_1(x)\phi_2(y)$ must contain infinitely
many conserved tensor fields. 

\smallskip

If all fields are scalars of dimension $2$, hence $\mu_{ij}\geq -1$,
double poles cannot occur in any $n$-point function subject to the
cluster decay property. We have exploited this fact in \cite{NRT2} to
prove that scalar fields $\phi$ of dimension $2$ are always Wick
products of the form $\sum M_{ij}\,\wick{\varphi_i\varphi_j}(x)$ of
massless free fields.  
In this argument, Hilbert space positivity plays a crucial role
because one has to solve a moment problem in order to get the correct
coefficients for all $n$-point functions simultaneously. (When we do
not insist that the theory possesses a stress-energy tensor with a
finite two-point function, then the fields $\phi$ may also have
contributions of generalized free fields.)

The simple pole structure of correlation functions of dimension 2
fields can be converted into commutation relations of the twist two 
biharmonic fields occurring in their OPE. The result is an
infinite-dimensional Lie algebra, whose unitary positive-energy
representations can be studied with methods of highest weight
modules. It turns out that there are no other representations than
those induced by the free field construction \cite{BNRT}.

\section{An example with double poles}
\setcounter{equation}{0} 

The following six-point structure solves the PDE (\ref{pde}) both
in the variables $x_1,x_2$ and in the variables $x_5,x_6$:
\bea\label{dp6}
u(x_1,\dots,x_6) =
  \frac{\left(\rho_{15}\rho_{26}\rho_{34} - 2\rho_{15}\rho_{23}\rho_{46} 
  - 2\rho_{15}\rho_{24}\rho_{36}\right)_{[1,2][5,6]}}
  {\rho_{13}\rho_{14}\rho_{23}\rho_{24}\cdot
  \rho_{34}^{d'-3}\cdot \rho_{35}\rho_{45}\rho_{36}\rho_{46}}\;,
\eea
where $(\dots)_{[i,j]}$ stands for the antisymmetrization in the arguments
$x_i$, $x_j$, and $\rho_{ij} = (x_i-x_j)^2$ as before. This structure
in addition obeys all homogeneity rules (\ref{sum}), pole bounds
(\ref{pb}) and cluster conditions in order to qualify as (a
contribution to) the correlation function  
\bea\label{U6}
\langle U(x_1,x_2)\phi'(x_3)\phi'(x_4)U(x_5,x_6)\rangle
\eea
where the scalar field $\phi'$ has dimension $d'$. The multiple
poles in the variables $x_3,x_4$ do not contradict the previous
argument (Sect.\ 3) excluding triple poles in the twist two ``channel'', 
when either $d$ (the dimension of the fields $\phi_1,\phi_2$ in (\ref{U}),
generating $U$) or $d'$ is $>2$, because they don't arise in a channel
of twist two ($1/\rho_{34}^{d'-3}$ is twist six, and $1/\rho_{3i}$ and 
$1/\rho_{4i}$ are twist two only if $d=d'=2$).

\smallskip

We determined the corresponding (contribution to the) correlation 
\bea\label{V6}
\vac{ V(x_1,x_2)\phi'(x_3)\phi'(x_4)V(x_5,x_6)},
\eea
$v(x_1,\dots,x_6)$, as the (simultaneous) harmonic part(s) of
$u(x_1,\dots,x_6)$. Let 
\bea\label{ccr}
s=\frac{\rho_{12}\rho_{34}}{\rho_{13}\rho_{24}}, \qquad
t=\frac{\rho_{14}\rho_{23}}{\rho_{13}\rho_{24}}
\eea
denote the conformal cross ratios, and $s'$ and $t'$ the same with
$1,2$ replaced by $5,6$. 
Then 
\bea\label{v6}
v(x_1,\dots,x_6)=u(x_1,\dots,x_6)\cdot g(t,s)g(t',s') + 
\hspace{20mm} \nonumber \\  +   
  \frac{2\left(\rho_{13}\rho_{24}\cdot\rho_{35}\rho_{46}\right)_{[1,2][5,6]}}
  {\rho_{13}\rho_{14}\rho_{23}\rho_{24}\cdot
  \rho_{34}^{d'-2}\cdot \rho_{35}\rho_{45}\rho_{36}\rho_{46}}\cdot
  (1-g(t,s)g(t',s'))
\eea
has the required power series expansion $u(x_1,\dots,x_6) +
O(\rho_{12},\rho_{56})$ provided $g(s,t)$ is of the form 
$g(s,t) = \sum_{n\geq 0} g_n(t)/n!\cdot s^n$ with $g_0(t)=1$, and it is
harmonic in all four variables $x_1,x_2,x_5,x_6$ provided $g$ solves
the PDE
\bea\label{diffg}
\Big((1-t\partial_t)(1+t\partial_t+s\partial_s) - 
[(1-t\partial_t)+t(2+t\partial_t+s\partial_s)]\partial_s\Big)\, g=0.
\eea
The solution is
\bea\label{g}
g(s,t) &=& \frac 1{s}\cdot\Big[Li_2(u)+Li_2(v) - Li_2(u+v-uv)\Big] + \\ 
 &+& \frac ts \cdot\left[Li_2\left(\frac {-u}{1-u}\right) 
 + Li_2\left(\frac {-v}{1-v}\right) 
 - Li_2\left(\frac{uv-u-v}{(1-u)(1-v)}\right)\right],\nonumber
\eea
where $u$ and $v$ (apologies for the duplicate use of letters!) here
stand for the ``chiral'' variables defined by the algebraic equations
\bea\label{uv}
s=uv\qquad\hbox{and}\qquad t = (1-u)(1-v).
\eea
$Li_2$ is the dilogarithmic function
defined by analytic continuation of its integral or power series
representations ($0\leq x < 1$)
\bea\label{dilog}
Li_2(x) = -\int_0^x\frac{\log(1-t)}t\,dt = \sum_{n>0}\frac{x^n}{n^2}.
\eea
Notice that $g$ is regular at $s=0$ in spite of the prefactors $\sim
1/s$. This transcendental correlation function can
definitely not be produced by free fields. 
It was found by turning the differential equation
(\ref{diffg}) into the recursive system  
\bea\label{recg}
(1+(n+1)t-t(1-t)\partial_t)g_{n} =
(1-t\partial_t)(n+t\partial_t)g_{n-1}
\eea 
with $g_0(t)=1$, 
and resumming the solution
\bea\label{gn}
\frac{g_n(t)}{n!} = \frac{n!(n+1)!}{(2n+1)!}\cdot{}_2F_1(n,n+1;2n+2;1-t)
\eea
by exploiting the integral representation of hypergeometric
functions.

\section{Local commutativity}
\setcounter{equation}{0} 

We shall now discuss the issue of local commutativity of the bi-field
$V(x,y)$. The naive argument would go as follows: since $U(x,y)$ is
Huygens bilocal in the sense of local commutativity for spacelike or
timelike separation from $x$ and $y$, the correlation functions  
\bea\label{uk}
u_k(x,y,\dots) = \vac{\phi_3(x_3)\dots\phi_k(x_k) U(x,y)
  \phi_{k+1}(x_{k+1})\dots\phi_n(x_n) } 
\eea
are independent of the position $k$ where $U(x,y)$ is inserted. By the
uniqueness of the harmonic decomposition, the same should be true for
their harmonic parts 
\bea\label{vk}
v_k(x,y,\dots) = \vac{\phi_3(x_3)\dots\phi_k(x_k) V(x,y)
  \phi_{k+1}(x_{k+1})\dots\phi_n(x_n) }, 
\eea
hence $V(x,y)$ commutes with $\phi_k(x_k)$.

However, this argument is not correct because of convergence problems
of the power series. The transcendentality of the correlation function
(\ref{v6}) shows that $V(x,y)$ in this case is certainly not a Huygens bilocal
field, which must have rational correlation functions by the same
argument \cite{NT} as for Huygens local fields. On the other hand,
the Result 2 in Sect.\ 3 yields a necessary and sufficient condition
(obviously violated by (\ref{dp6})):

\smallskip

\noindent 
\R3{$V(x,y)$ is Huygens bilocal if and only if the coefficients of
the twist two pole $((x-y)^2)^{-(d-1)}$ in every correlation involving
$\phi(x)\phi(y)$ (i.e., the leading parts $u_0$ of the correlations
of $U(x,y)$) never exhibit ``double poles'' in the variables $x$ or
$y$ (as explained Sect.\ 3).  }

\smallskip

In general, i.e., when there are double poles, $V(x,y)$ is originally
only defined as a formal power series (in $x-y$) within each
correlation function. Even when these series converge, it is not a
priori clear what the labelling pair of points $x,y$ has to do with
its localization in the sense of local commutativity with other
fields, because splitting the OPE into pieces is a highly nonlocal
operation (involving projections onto eigenspaces of conformal Casimir 
operators).    

\smallskip 

In order to study local commutativity, we need to control convergence
of the series defining the harmonic part. The latter can be addressed
with the help of the ``generalized residue formula''. This integral
representation of the harmonic part was found recently \cite{BN} in
the context of higher-dimensional vertex algebras:
\bea\label{res}
v(x)=\frac1{i\pi\vert\SS^{D-1}\vert}
\int_{M_{r}}d^Dz\vert_{M_r}\frac{1-x^2/z^2}{((z-x)^2)^{D/2}}\; u(z). 
\eea
Here, $z^2=\sum_{a=1}^D z_a^2$ is the complex Euclidean square. $M_r$
is the compact submanifold $r\cdot \SS^1\cdot \SS^{D-1}$
($\SS^1\subset\CC$ is the complex unit circle, and
$\SS^{D-1}\subset\RR^D\subset\CC^D$ the real unit sphere), and
$d^Dz\vert_{M_r}$ the induced complex measure.  

The radius $r>0$ has to be chosen such that $u(z)$ converges absolutely
for $z\in M_r$. Then for $x$ small enough such that the kernel
converges absolutely as a power series in $x$ for every $z\in M_r$,
the integral converges as a power series in $x$, and is independent of
the choice of $r$. 

This formula for the harmonic part w.r.t.\ the Euclidean Laplacian 
remains valid for the Lorentzian Laplacian, provided $z^2$ is replaced
by the (complex) Lorentzian square, and the unit sphere by the set
$\{(ix^0,\vec x): (x^0,\vec x\in\SS^{D-1})\}$. This is true because
the map $\CC^D\to\CC^D$, $(z^0,\vec z)\mapsto (iz^0,\vec z)$,
intertwines the Euclidean with the Lorentzian harmonic decomposition. 

\smallskip 

In the case at hand, where $x-y$ plays the role of $x$ and $u(x)$ is
the Taylor series around $x$ or around $y$, respectively, of a Laurent
polynomial with poles at $(x-x_j)^2=0$ and $(y-x_j)^2=0$, we find
absolute convergence in the domain
\bea\label{dom}
\norm{x-y} + \sqrt{\norm{x-y}^2+\vert(x-y)^2\vert} < \hspace{37mm} 
\nonumber \\ < 
\sqrt{\norm{x-x_j}^2+\vert(x-x_j)^2\vert} - \norm{x-x_j}
\quad\forall\;j=1,\dots n
\eea
in the first case, and the same with $x$ replaced by $y$ in the
second case. Especially, if $x$ and $y$ are spacelike or timelike
separated from all other points $x_j$, these domains are not empty. 
We have therefore

\smallskip

\noindent 
\R4{The formal power series $v_k(x,y,\dots)$ for the 
correlation functions (\ref{vk}) converge absolutely within the
domains (\ref{dom}), and the resulting functions $v_k$ do not depend
on the position $k$ where $V(x,y)$ is inserted in (\ref{vk}).}

\smallskip 

The issue of local commutativity of $V(x,y)$ with $\phi_k(x_k)$ now
amounts to the question whether $v_{k-1}=v_{k}$ still holds outside
the domain (\ref{dom}), as long as $x_k$ is spacelike (or timelike)
from $x$ and $y$. We conservatively anticipate that the correlation
functions are real analytic functions of real spacetime points within
the region where local commutativity holds. Then the existence of 
a unique real analytic continuation from (\ref{dom}) to some other
configuration implies $v_{k-1}=v_{k}$ at the latter configuration by
virtue of Result 4, and hence commutativity. Continuation beyond a
singularity requires to go through a suitable complex cone which 
depends on the position $k$ where $V(x,y)$ is inserted in (\ref{vk}),
hence commutativity will fail.
Put differently, our strategy to establish locality by inspection of
analyticity inverts the usual axiomatic reasoning \cite{SW} by which
one {\em derives} the domain of analyticity from the {\em known} locality
(and energy positivity). 

\smallskip 

We want to discuss specifically the local commutativity of
$V(x_1,x_2)$ with $\phi'(x_3)$ in the case of the example (\ref{v6}),
by studying its maximal real analytic continuation starting from the
domain (\ref{dom}), which is a neighborhood of $x_1=x_2$ where $s=0$,
$t=1$, hence $u=v=0$. Clearly, we can only reach configurations where
$(x_1-x_k)^2\neq 0$ has the same sign as $(x_2-x_k)^2$ for $k=3$ and
$k=4$, because this is trivially true at $x_1=x_2$ and we cannot pass
through $t=0$ or $t=\infty$ where the variables $u$ or $v$ in
(\ref{g}) would hit the singularities of the dilogarithmic function
$Li_2(z)$ at $z=1$ and $z=\infty$. 

We claim that (\ref{v6}) has a
unique real analytic continuation to {\em all} these points, or
equivalently, that $g(s,t)$ given by (\ref{g}) has a unique real
analytic continuation in the region $t>0$, $s$ arbitrary (real). This
is obvious for the last terms in the two lines of (\ref{g}) because
for $t>0$ their arguments are $<1$. For the study of the remaining
terms, we solve (\ref{uv}) for $u$ and $v$ (where it does not matter
which one is which because of the manifest symmetry of (\ref{g}) under
$u\leftrightarrow v$)  
\bea\label{chi}
u,v = \frac12\Big(1-t+s\pm\sqrt{(1-t+s)^2-4s}\Big). 
\eea
In the range $s \leq
(1-\sqrt t)^2$, $u$ and $v$ are real and $u+v = 1-t+s<2$. From
$(1-u)(1-v)=t>0$, we see that both $u$ and $v$ are $<1$, and so are
$\frac{-u}{1-u}$ and $\frac{-v}{1-v}$.   
The continuation to these points is unambiguous. In the range
$(1-\sqrt t)^2<s<(1+\sqrt t)^2$, $u$ and $v$ are complex and conjugate
to each other, so that the first two terms in both lines of (\ref{g})
are always the sum of the values on the two branches above and below
the cut. In particular, $g(s,t)$ is real and $Li_2$ in (\ref{g}) may
be replaced by its real part. Finally, in the range $s \geq (1+\sqrt
t)^2$, we find $u+v >2$, hence both $u$ and $v$ and also
$\frac{-u}{1-u}$ and $\frac{-v}{1-v}$ are $>1$. All four arguments
hit the cut of $Li_2(z)$. But because its discontinuity is 
{\em imaginary}, the real parts are real analytic. This proves the
claim. 

As explained before, the maximal domain of real analyticity specifies
those configurations $x_1,x_2,x_3$, where $\phi'(x_3)$ commutes with
$V(x_1,x_2)$. We may assume $x_4^2\to\pm\infty$ (which can be achieved
by a conformal transformation), hence
$t=(x_2-x_3)^2/(x_1-x_3)^2$. Thus we get commutativity whenever $x_3$
is {\em simultaneously} spacelike or timelike from $x_1$ {\em and} $x_2$. 
We summarize

\smallskip

\noindent 
\R5{The transcendental (part of a) correlation function
(\ref{v6}) is compatible with local
commutativity between $V(x,y)$ and $\phi'(z)$ when $x-z$ and
$y-z$ are either both spacelike or both timelike.}

\smallskip

The set of these configurations is locally, but {\em not globally}
conformal invariant, since a conformal transformation may switch the
sign $\sigma$ of $(x-z)^2/(y-z)^2$. This is not a
contradiction: connecting configurations with $\sigma=+$ with those
with $\sigma=-$ by a path in the conformal group, one must necessarily
pass through $x=\infty$ or $y=\infty$, where the OPE in terms of power
series ceases to make sense. This breakdown of GCI for the biharmonic
field $V(x,y)$ is, of course, just another manifestation of its
violation of Huygens bilocality.   

\smallskip

It is worth noticing that another decomposition theory for the
OPE in conformal QFT was developed in \cite{SS,SSV}. While it
is coarser than the twist decomposition (it is even trivial in the 
GCI case), it was found to exhibit, at least in two dimensions
\cite{RS}, a ``localization between the points'' with similar
implications as the present one. 

\section{Conclusion}
We have outlined recent progress in the intrinsic structure analysis
of quantum field theories in four dimensions, under the assumption of
``global conformal invariance'' \cite{NRT2}.

We have found nontrivial restrictions on the singularity structure of
correlation functions. The encouraging aspect is that these restrictions
allow a small ``margin'' beyond free correlations, for which we have
given a nontrivial example. It exhibits a local but not Huygens local
bi-field $V(x,y)$ whose correlation functions involve dilogarithmic
functions. Local commutativity with a third field at a point $z$ is
shown to hold (in this example) whenever $x-z$ and $y-z$ are
either both spacelike or both timelike. The possible failure of local
commutativity when one is spacelike while the other is timelike,
occurs only in correlations of at least five points, because
four-point functions cannot exhibit the characteristic ``double
poles'' in the twist two channel that are responsible for the
transcendental correlations involving $V(x,y)$.  

A serious question remains to be answered before our six-point
structure is established as (a contribution to) a manifestly non-free
correlation function: we cannot control (at the moment) Wightman
positivity at the six-point level. In the case at hand, this means
that we do not know whether the vectors $\phi'(z)V(x,y)\vert 0\rangle$
span a Hilbert (sub-)space with positive metric. Because our six-point
structure reduces in the leading OPE channels to five- and four-point
structures that can also be obtained from free fields, positivity can
only be violated in higher channels where our present knowledge of
partial waves is not sufficient. Far more ambitious is the problem
whether a given six-point function can be supplemented by higher
correlations  satisfying Wightman positivity in full generality, i.e.,
to recover the full Hilbert space on which the bi-field $V(x,y)$ and
its generating field $\phi(x)$ act. 

%


\section*{Acknowledgments}

The authors thank the organizers of the conference ``LT7 -- Lie
Theory and its Applications in Physics'' (Varna, June 2007) for giving
them the opportunity to present these results, and the
Alexander von Humboldt Foundation for financial support.

\end{document}